\documentclass[twocolumn,aps,showpacs,prl,amsmath,amssymb,floatfix,superscriptaddress]{revtex4-1}
\usepackage{color}
\usepackage{graphicx}
\usepackage{dcolumn}
\usepackage{bm}
\usepackage{array}
\usepackage{float}
\usepackage{supertabular}
\usepackage{longtable}
\begin{document}

\newcommand{\lprime}{l\hspace{-.3mm}'}

\title{How to find conductance tensors of quantum
multi-wire junctions through static calculations: application to an interacting Y junction}

\author{Armin Rahmani}
\affiliation{Department of Physics, Boston University, Boston,
MA 02215 USA }
\author{Chang-Yu Hou}
\affiliation{Department of Physics, Boston University, Boston, Massachusetts 02215 USA }
\affiliation{Instituut-Lorentz, Universiteit Leiden, P.O. Box 9506, 2300 RA Leiden, The Netherlands}
\author{Adrian Feiguin}
\affiliation{Department of Physics and Astronomy, University of Wyoming, Laramie, Wyoming 82071, USA}
\author{Claudio Chamon}
\affiliation{Department of Physics, Boston University, Boston,
MA 02215 USA }
\author{Ian Affleck}
\affiliation{Department of Physics and Astronomy, University of British
Columbia, Vancouver, British Columbia, Canada, V6T 1Z1}

\date{\today}

\begin{abstract}

Conductance is related to dynamical correlation functions which can be calculated with \textit{time-dependent} methods. Using boundary conformal field theory, we relate the conductance tensors of quantum junctions of multiple wires to static correlation functions in a finite system. We then propose a general method for determining the conductance through \textit{time-independent} calculations alone. Applying the method to a Y junction of interacting quantum wires, we numerically verify the theoretical prediction for the conductance of the chiral fixed point of the Y junction and then calculate the thus far unknown conductance of its M fixed point with the time-independent density matrix renormalization group method. 

\end{abstract}

\maketitle

Advances in molecular electronics can extend the limits of device
miniaturization to the atomic scales where entire electronic circuits
are made with molecular building blocks~\cite{Nitzan03}. Single
molecule junctions connected to two macroscopic metallic leads have
already been successfully fabricated~\cite{Reed97}. A key ingredient
of any such molecular circuit is a junction of three or more
quantum wires where electrical current is conducted through a
molecular structure between the wires. Determining the conductance of
such junctions is a long-sought and challenging goal.

Landauer-B\" uttiker's formalism does not account for electron-electron
interactions which play a key role in low dimensions. Functional renormalization group methods have been helpful in studying the interaction effects~\cite{Barnabe05} but they are also dependent upon the presence of noninteracting leads. There are a number of challenging problems in junctions of multiple interacting wires. For instance, a time-reversal symmetric Y junction of Luttinger
liquids has a nontrivial fixed point known as the M fixed point~\cite{Chamon06}, which has remained an open quantum impurity problem. For spinfull electrons, there are nontrivial fixed points even for junctions of two quantum wires~\cite{Kane92}. Numerical methods like density matrix renormalization group (DMRG)~\cite{White92,Guo06} could potentially be efficient tools for computing the conductance of junctions with an arbitrary number of wires and interactions. Conductance is a property of an open quantum system and is related to
dynamical correlation functions. It may thus appear that one needs the more computationally demanding dynamical methods such as time-dependent DMRG for conductance calculations~\cite{Marston,Luo,White-Feiguin-Daley,Schmitteckert,Al-Hassanieh,Heidrich-Meisner,Silva}. These require large systems and averaging over a time window.


The goal of this Letter is to find generic relations that permit the
computation of conductance from static equilibrium calculations
alone, thus making it possible to obtain the conductance tensors of
arbitrary complex junctions with any number of leads using 
time-independent methods, such as standard DMRG. As a concrete application, we are interested in the open problem of the conductance of the M point of a Y junction of quantum wires. 
We argue that the $T=0$ linear conductance $G_{ij}$ (defined 
by $I_i=\sum_j G_{ij}V_j$ where $V_j$ is the voltage applied to wire $j$ 
and $I_j$ is the inward directed current on it) is determined by the 
conformally invariant boundary condition (BC) describing the infrared 
fixed point of quantum junctions. Finding the conductance could then be 
helpful in determining nontrivial fixed points.

Using boundary conformal field theory, we find a generic relationship
between the \textit{dynamical} correlation functions of a \textit{semi-infinite} quantum
junction and the \textit{static} correlation functions of a \textit{finite} system obtained by a conformal transformation. This relationship
allows us to extract certain coefficients from a time-independent numerical
calculation of the ground state expectation values of appropriate local operators which
uniquely determine the dynamical correlation functions and give us the linear-response
conductance of the junction through the Kubo formula. After
establishing the generic continuum formalism, we discuss the application of the formalism to a discrete lattice computation.

We present DMRG results on the application of the method to an interacting Y junction of quantum wires. We successfully verify the theoretical prediction~\cite{Chamon06} of $G_{12}=- 2\:\frac{g}{3+g^2}(g+1)\frac{e^2}{h}$ for the chiral fixed point of a Y junction and numerically calculate the conductance at two values of the Luttinger parameter $g$ for the time-reversal symmetric M fixed point.

\begin{figure}
\includegraphics[width=8.5 cm]{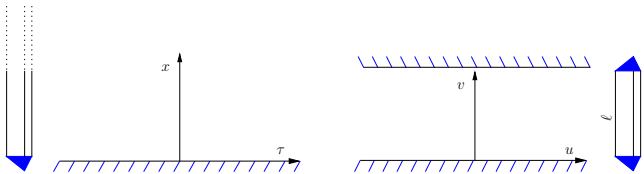}
\caption{The conformal mapping from the upper half-plane to the strip and the corresponding physical systems. The actual junction (here a Y junction, for instance) is illustrated next to to the corresponding complex plane regions. The left panel shows the semi-infinite junction (the upper half-plane) and the right panel shows the finite system obtained by the conformal mapping (the strip). }
\label{fig_map}
\end{figure}

Consider $M$ semi-infinite wires connected to a junction with
arbitrary structure and interactions. We represent the
system as a quasi-one-dimensional structure with all the wires running
parallel to the positive $x$ axis (Fig.~\ref{fig_map} left panel). The
wires are connected to the molecular structure of the junction at
$x=0$. Setting the velocity of the charge carriers to unity, we introduce the complex coordinates $z=\tau+i x$, where $\tau$
is the imaginary time. The multiple wires are described by $M$ species
of electrons living on the upper half complex plane. Notice that spin
can also be taken into account by doubling the number of species, but here we focus on spinless electrons.

The low energy behavior of the system is given by a renormalization
group fixed point. Here we treat the junction problem as a boundary conformal field theory~\cite{Cardy89}. In the bulk of the wires, we have a conformal field
theory (CFT) with central charge $M$ and bosonic fields $\phi_j, \;
j=1, \dots, M$ obtained from the fermionic fields via standard
bosonization. We assume that the effect of the junction (the molecular
structure connecting the wires) on the fixed-point behavior is
generically encoded in a {\it conformally invariant} boundary
condition for the CFT in the upper half-plane and represented by a
boundary state $|{\cal B}\rangle$~\cite{Wong94}. This assumption is at the
heart of a paradigm in quantum impurity problems and has been
repeatedly verified in a multitude of such problems~\cite{Affleck08}.

Let us first consider the CFT in the absence of the boundary. In terms
of the bosonic fields $\phi_j$ or their dual fields $\theta_j$, the
action for this system is given by
\[S=\sum_j {g \over 4 \pi} \int d^2 x \:\partial _\mu \phi^j \partial^\mu \phi^j=\sum_j {1 \over 4 \pi g} \int d^2 x \:\partial _\mu \theta^j \partial^\mu \theta^j\]
where $g$ is the Luttinger parameter. For noninteracting wires $g=1$
and we have $g<1$ ($g>1$) for repulsive (attractive) interactions.
The primary operators for this CFT are the vertex operators $e^{i
\alpha \phi_j}$ (the fermion creation and annihilation operators are
of this form) and the currents
\[J^j_L(z)=\frac{i}{\sqrt{2}\pi}\: \partial \: \theta^j(z,\bar{z})\, \qquad J^j_R(\bar{z})=-\frac{i}{\sqrt{2}\pi}\: \bar{\partial} \: \theta^j(z,\bar{z})\]
where $\partial \equiv \partial_z={1 \over
2}(\partial_\tau-i\partial_x)$ and $\bar{\partial} \equiv
\partial_{\bar {z}}={1 \over 2}(\partial_\tau+i\partial_x)$. In the absence of a boundary, the only nonvanishing correlation functions are the chiral ones: 
\begin{equation}\label{chiral_corr}
\langle {\cal T}_\tau J_L^i(z_1)J_L^j(z_2)\rangle= -{g\over 4 \pi ^2}\frac{\delta_{ij}}{(z_1-z_2)^2}
\end{equation}
and similarly for $\langle {\cal T}_\tau J_R^i(\bar{z}_1)J_R^j(\bar{z}_2)\rangle$ where ${\cal T}_\tau$ indicates imaginary time-ordering.

Let us now consider the system on the upper half-plane. In general,
the presence of the boundary does not change the correlation between
primary operators of the same chirality~\cite{Affleck93} but
introduces additional correlations between the left movers and the
right movers,
\begin{equation*}
  \langle {\cal T}_\tau J_L^i(z_1)J_R^j(\bar{z}_2)\rangle= -{g\over 4 \pi ^2}A_{\cal B}^{ij}\frac{1}{(z_1-\bar{z}_2)^2} \; . 
\end{equation*}
The coefficients $A_{\cal B}^{ij}$ are determined by the BC on the real axis.  
In fact, in terms of the boundary state $|{\cal B} \rangle$, we have
$A_{\cal B}^{ij}=\frac{\langle J_L^iJ_R^j,0|{\cal
B}\rangle}{\langle\openone,0|{\cal B}\rangle}$ where $|O,0\rangle$
is the highest weight state corresponding to a generic operator $O$ (here $O=J^i_L J^j_R$)~\cite{Cardy92}. Note that at the fixed point
the correlators are determined by conformal symmetry up to the
coefficients above and as long as we are interested in the fixed-point
conductance of a given junction, a direct numerical calculation of
dynamical correlation functions is not required. The fact that the
dynamical correlations are directly related to equal-time correlations
by conformal invariance is the {\it key reason} why we can obtain the
conductance from a time-independent calculation.

If we know the coefficients $A_{\cal B}^{ij}$, we can
use the Kubo formula~\cite{Chamon06}
\[
G_{ij} = \lim_{{\omega} \rightarrow 0_+} -\frac{e^2}{\hbar}\frac{1}{
	{\omega} L} \int_{-\infty}^\infty \!\!\!\! d\tau\;e^{i {\omega}\tau}
	\!\! \int_0^L \!\!\!\! dx \;\langle {\cal T}_\tau J^i(y,\tau) J^j(x,0)\rangle
\]
with $J^j=J^j_R-J^j_L$ to obtain the
conductance of the junction upon performing the integrations.
Let us focus on calculating the off-diagonal elements of the
conductance tensor $G_{ij}$. The diagonal elements are not independent
since $\sum_{i}G_{ij}=\sum_{j}G_{ij}=0$ due to current conservation
and the fact that a common voltage applied to all wires gives zero
current.  For $i\neq j$, as seen in Eq.~(\ref{chiral_corr}), the
correlators of the same chirality vanish.  Using the Kubo formula
above and after doing the $\tau$ integral by contour integration we
obtain for $i\neq j$
\begin{eqnarray}\label{Kubo}
G_{ij} &=& g \frac{e^2}{ h}\frac{1}{L}
	 \int_0^L dx [A^{ij}_{\cal B} \;H(x+y)+A^{ji}_{\cal B}\; H(-x-y)]
\nonumber\\
&=& A^{ij}_{\cal B} \; g\frac{e^2}{ h}
\end{eqnarray}
where $H$ is the Heaviside step function. Since both $x$ and $y$ are positive, $H(-x-y)=0$ in the second term.

The next task is to extract $A^{ij}_{\cal B}$ by measuring an
equal-time ground state correlation function. The conductance is
well-defined for a semi-infinite system, while the equilibrium
equal-time correlators are easily obtainable in a finite system. To
get around this issue, we use the following conformal
transformation to map the upper half-plane $z=\tau + ix$ into a strip
of width $\ell$,
\begin{equation}\label{mapping}
w={\ell \over \pi} \ln z, \qquad z=e^{{\pi \over \ell }w}.
\end{equation}
Notice that for $w=u+i v$ we have $0<v<\ell$. The mapping takes the
positive real axis to the boundary $v=0$ and the negative real axis to
$v=\ell$.

Now consider a primary operator $O$ (here $O=J_L^iJ_R^j$) with
\[\langle O(z) \rangle=A_{\cal B}^O(2x)^{-X_O}\]
in the semi-infinite plane with BC ${\cal B}$ on the real
 axis. Using the transformation $\langle O(w) \rangle=|{dw \over dz}|^{-X_O}\langle O(z) \rangle$ we obtain after some algebra
\[\langle O(w) \rangle=A_{\cal B}^O \left[2\:{\sin\left( {\pi \over \ell}v\right) }/{{\pi \over \ell}} \right]^{-X_O} \]
Since the zero temperature finite system is invariant under
translations in $u$, the static ground state expectation value of the
local operator $J_L^i(x)J_R^j(x)$ with $x$ the distance from a
boundary in the finite system of length $\ell$ is expected to behave as $-{g\over 4 \pi ^2}A_{\cal B}^{ij} \left[2\:
{\sin\left( {\pi \over \ell}x\right) }/{{\pi \over \ell}}
\right]^{-2}$. This ground state expectation value permits the determination of the $A_{\cal B}^{ij}$ coefficients and
thus the conductance via Eq.~(\ref{Kubo}) using solely static
equilibrium computations.

In the remainder of this Letter, we discuss the application of the method to a tight-binding
lattice numerical calculation. First we consider the BCs for the finite system obtained from the transformation Eq.~(\ref{mapping}).
Because we are assuming a conformally invariant BC, the boundary $v=0$ will have the same BC ${\cal
B}$ as the real axis of the semi-infinite plane. To determine the BC for the $v=\ell$
boundary created by the transformation, we consider the fermion creation and annihilation operators which transform as
\begin{equation*}
\Psi_L(w)=\left(\frac{dw}{dz} \right)^{-\frac{1}{2}}\Psi_L(z), \quad \Psi_R(\bar{w})=\left(\frac{d\bar{w}}{d\bar{z}} \right)^{-\frac{1}{2}}\Psi_R(\bar{z}).
\end{equation*}
Using $\frac{d\bar{w}}{d\bar{z}}=\frac{dw}{dz}{\big |}_{v=0}=\frac{\ell}{\pi}e^{-\frac{\pi}{\ell}u}$ and $\frac{d\bar{w}}{d\bar{z}}=\frac{dw}{dz}{\big |}_{v=\ell}=-\frac{\ell}{\pi}e^{-\frac{\pi}{\ell}u}$, we find that up to a sign coming from choosing the branch of the square root in the transformation above, the BC at $v=\ell$ is the same as $v=0$. This does not mean however that in a microscopic implementation, one should place the mirror image of the junction at $v=\ell$. Notice that the chiral fields switch role for the two boundaries; i.e. the right-movers are the outgoing (incoming) states for the  $v=0$ ($v=\ell$) boundary. So for instance if the BC at $v=0$ is described by an S-matrix $S$, the BC at $v=\ell$ is given by the inverse S-matrix $S^\dagger$ up to a sign.

We consider a half-filled system with an even number of sites in each wire and particle-hole symmetry in the bulk of the wires. The Hamiltonian of the system is given by $H_L+H_B+H_R$ where $H_B$ is the bulk Hamiltonian of the wires and $H_{L,R}$ are the junction Hamiltonians. In terms of the fermionic operators $\Psi^j_m$ on lead $j$ and site $m=1,\dots l$, we have
\[
H_B=\sum_{j,m}\left[-\Psi^{j\dagger}_m \Psi^j_{m+1}+{\rm h.c.}+V (n^j_m-\frac{1}{2}) (n^j_{m+1}-\frac{1}{2}) \right]
\]
where $n^j_m\equiv{\Psi^j}^\dagger_{m}\Psi^j_m$, the hopping amplitude is set to unity and $V$ is the interaction strength. To implement the correct boundary condition at $v=\ell$, we construct $H_R$ by applying a particle-hole and a time-reversal transformation to $H_L$. As a concrete example we have the boundary contribution $H_R=-\Psi^{i \dagger}_l \Gamma_{i,j}\Psi_l^j$ if $H_L=\Psi^{i \dagger}_1\Gamma_{i,j}\Psi_1^j$ for a Hermitian matrix $\Gamma$.

The coefficients $A_{\cal B}^{ij}$ can be extracted from the ground state correlators
of \textit{chiral} operators in a finite system. We cannot directly
model chiral fermionic creation and annihilation operators, but indeed
we can model chiral currents using density and current
operators. The (nonchiral) current operator
on the link between sites $m$ and $m+1$ is given by
$J^j_m=i({\Psi^j}^\dagger_{m+1}\Psi^j_m-{\Psi^j}^\dagger_{m}\Psi^j_{m+1})$. Also, for a bond between sites $m$ and $m+1$, the charge density operator (with the background charge subtracted) is given by
$N^j_m={1 \over 2}\left (n^j_m+n^j_{m+1}-\langle n^j_m \rangle-\langle n^j_{m+1}\rangle\right).$ Chiral current operators on the lattice are related to the density and nonchiral current through
\begin{equation}\label{J and N}
 J^j(x)=v\:(J^j_R(x)-J^j_L(x)), \quad N^j(x)=J^j_R(x)+J^j_L(x)
\end{equation}
where $v$ is the velocity of charge carriers, and $x$ takes values in the middle of the bonds, $x=m+1/2$.

The Luttinger parameter $g$ and the velocity $v$ depend on the strength of the interaction~\cite{Giamarchi03,Imura02}. We have from the Bethe ansatz
\begin{equation}
g=\frac{\pi}{2\arccos\:(-V/2)},\qquad
v=\pi\frac{\sqrt{1-(V/2)^2}}{\arccos\:(V/2)}.
\end{equation}

We would like to find the off-diagonal elements of the conductance tensor $G_{ij}$. The
chiral correlators are proportional to $\delta_{ij}$ in the half-plane and therefore vanish for $i\neq j$. They also vanish in the finite system since it is obtained from the half-plane by a conformal mapping. Using Eq.~(\ref{J and N}), we can then
write for $i\neq j$
\begin{eqnarray}
  \langle J^i(x)J^j(x)\rangle&=&-v^2\left(  \langle J_L^i(x)J_R^j(x)\rangle+\langle J_R^i(x)J_L^j(x)\rangle\right),\nonumber \\ 
 \langle N^i(x)J^j(x)\rangle&=&v\left(  \langle J_L^i(x)J_R^j(x)\rangle-\langle J_R^i(x)J_L^j(x)\rangle\right) .
\end{eqnarray}
By measuring the above two correlation functions (the second correlator $\langle N^i(x)J^j(x)\rangle$ is identically zero in the absence of a magnetic flux), we can find the chiral correlators we need for calculating the conductance. 
We fit the measured $\langle J_L^i(x) J_R^j(x)\rangle$ with $\tilde{A} \left[2\: {\sin\left( {\pi
\over \ell}x\right) }/{{\pi \over \ell}} \right]^{-2}$ to find the coefficient $\tilde{A}$ and then obtain the conductance using Eq.~\ref{Kubo}. Note that the fixed point behavior cannot be observed close to the boundary so the numerical results for small $x$ must be excluded in fitting the data.

\begin{figure}
 \centering
 \includegraphics[width=8 cm]{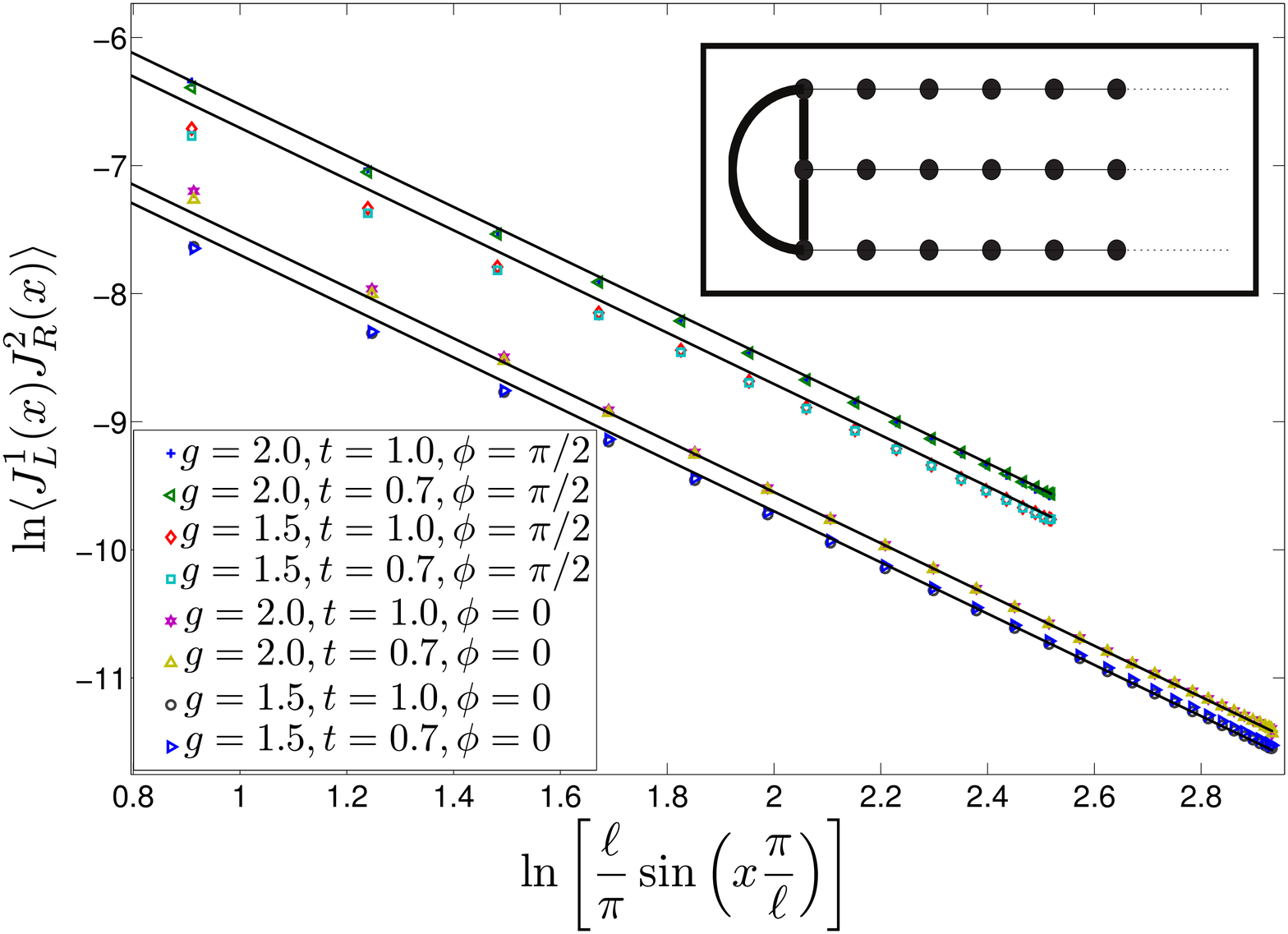}
 \caption{The inset shows the lattice Y junction. The thick bonds at the junction have a  hopping amplitude $t$. The lines drawn through the data points show the CFT prediction with the exact analytical conductance in the case of the chiral fixed point ($\phi\neq0$) and the fit used for obtaining the conductance in the case of the M fixed point ($\phi=0$). As expected, the conductance is independent of $t$.}
 \label{fig_corr}
\end{figure}

Below we consider an interacting Y junction of three quantum wires. As a check,
we first apply the method to the chiral fixed point whose conductance has been predicted in Ref.~\cite{Chamon06}. We then present the first numerical calculation of the conductance of the time-reversal symmetric M fixed point whose properties have so far remained unknown. As seen in the inset of Fig.~\ref{fig_corr}, the junction simply connects the three endpoints of the wires with a hopping amplitude $t$ (the amplitude is $1$ in the bulk). Assuming that the loop
formed at the junction is threaded by a magnetic flux $\phi$, the system will renormalize to the chiral fixed point for nonzero $\phi$ and $1<g<3$ independently of the hopping amplitude $t$. In the same range of $g$, the system renormalizes to the M fixed point in the absence of a magnetic flux ($\phi=0$). 

The numerical calculations for both fixed points were performed for $g=1.5, 2.0$ and $t=0.7, 1.0$. For the chiral fixed point, using $\phi=\pi/2$, $\ell=39$ ($40$ sites on each chain) and the number of DMRG states $m=600$, we found strong agreement between the numerics and the predicted conductance $G_{12}=- 2\:\frac{g}{3+g^2}(g+1)\frac{e^2}{h}$~\cite{Chamon06} as seen in Fig.~\ref{fig_corr}. The correlation functions are independent of the hopping amplitude $t$ indicating convergence to the universal conductance of the fixed point. For the M fixed point, using $\phi=0$, $\ell=59$ and $m=1000$, we obtain a conductance of $G_{12}=-0.5293\frac{e^2}{h}$ ($G_{12}=-0.6146\frac{e^2}{h}$) for $g=1.5$ ($g=2.0$). None of the properties of the M fixed point were previously known in the literature~\cite{Affleck08}. These values of the conductance depend on $g$ nontrivially and are different from those of all the other fixed points which have been identified so far. Thus our method and the numerical results provide evidence that the M fixed point is of a different nature.

In summary, we presented in this Letter a generic relationship between
the elements of the conductance tensor of quantum junctions at a
renormalization group fixed point and coefficients in the
static correlation functions of chiral current operators. The method
hinges on a fundamental relationship between static and dynamic
correlation functions in boundary conformal field theory. We proposed
a method for obtaining conductances from static calculations by
extracting these coefficients using nonchiral current and density correlations in a lattice computation. We applied the method to an interacting Y junction of Luttinger liquids using time-independent DMRG and presented a numerical verification of the theoretically predicted conductance of the chiral fixed point as well as a calculation of the previously unknown conductance of the M fixed point. We emphasize that this general method turns the time-independent DMRG into a powerful tool for extracting conductances of completely generic molecular electronic junctions which constitute a wide class of quantum impurity problems.

\acknowledgments{We are grateful to A. Sandvik for
helpful discussions. This work was supported in part by the DOE Grant
No. DE-FG02-06ER46316 (CC, CH, and AR), NSF DMR-0955707 (AF), NSERC (IA) and CIfAR (IA).}

\end{document}